\documentclass[
reprint,
amsmath,amssymb,
aps,
prl,
longbibliography]{revtex4-2}

\usepackage{xspace}
\usepackage{graphicx}
\usepackage{bm}
\usepackage[colorlinks=true,urlcolor=blue,linkcolor=blue,citecolor=blue,bookmarks=false]{hyperref}
\usepackage{amsmath}
\usepackage{amssymb}
\usepackage{bbold}
\usepackage{soul}
\usepackage{dsfont}
\usepackage{braket}
\usepackage{textcomp}

\newcommand{\customref}[2]{\hyperref[#1]{\ref*{#1}#2}}

\usepackage{xcolor}

\definecolor{Ured}{HTML}{cc0000}
\definecolor{Ublue}{HTML}{1f65cf}
\definecolor{Ugreen}{HTML}{198a11}

\newcommand{\ie}[0]{i.e.\@\xspace}
\newcommand{\eg}[0]{e.g.\@\xspace}

\newcommand{\cf}[0]{cf.\@\xspace}

\newfont{\tensy}{cmsy10}

\renewcommand{\S}[0]{\mathcal{S}}

\newcommand{\fden}[1]{\hat{n}_{#1}}

\newcommand{\Q}[1]{\hat{Q}_{#1}}
\renewcommand{\P}[1]{\hat{P}_{#1}}

\newcommand{\fcohan}[1]{c_{#1}}
\newcommand{\fcohcr}[1]{\bar{c}_{#1}}
\newcommand{\fcohmeasure}{\mathcal{D}(\bar{c},c)}

\newcommand{\kF}{k_{\text{F}}}

\newcommand{\Hc}{\mathrm{H.c.}}

\newcommand{\expv}[1]{\left\langle #1 \right\rangle}

\newcommand{\gb}{g_\mathrm{b}}

\begin{document}
\title{Competing Dirac masses in one dimension:\\Symmetry-enhanced pseudo-first-order transition and deconfined criticality}

\author{Manuel Weber}
\affiliation{Institut f\"ur Theoretische Physik and W\"urzburg-Dresden Cluster of Excellence ct.qmat, Technische Universit\"at Dresden, 01062 Dresden, Germany}

\date{\today}

\begin{abstract}
Emergent symmetries and slow crossover phenomena are central themes in quantum criticality and manifest themselves in the pseudocritical scaling experienced in the context of deconfined criticality. Here we discover its conceptual counterpart, i.e., a symmetry-enhanced \textit{pseudo-first-order transition}. It emerges from a one-dimensional realization of deconfined criticality between charge- and bond-ordered states driven by competing Holstein and Su-Schrieffer-Heeger electron-phonon couplings, for which quantum fluctuations and thereby the nature of the transition can be tuned systematically via the phonon frequency $\omega_0$. In the classical limit $\omega_0 \to 0$, a low-energy Dirac theory predicts a direct first-order transition with emergent U(1) symmetry. Using exact quantum Monte Carlo simulations, we provide strong evidence for symmetry enhancement and even finite-size scaling on intermediate length scales but in the thermodynamic limit it turns into a narrow intermediate phase where both order parameters are finite, as chiral U(1) symmetry is weakly broken on the lattice. Including quantum lattice fluctuations diminishes the width of the intermediate phase, gradually restores the U(1) symmetry, and eventually tunes the system to a deconfined quantum critical point.
\end{abstract}

\maketitle

\textit{Introduction.}---%
Competing orders are a central theme in quantum materials as they can lead to complex phase diagrams and exotic criticality.
\textit{Deconfined quantum criticality} is a novel mechanism in which fractionalized excitations turn a direct order-to-order transition with distinct broken symmetries continuous, a scenario that is forbidden within Landau theory \cite{Senthil1490, PhysRevB.70.144407}. First proposed for two-dimensional SU(2) quantum magnets \cite{PhysRevLett.98.227202}, this transition
 is now considered weakly first-order \cite{senthil2023deconfined}. 
The search for deconfined criticality has brought forth a variety of closely-related phenomena like symmetry enhancement at continuous \cite{PhysRevLett.115.267203} and even first-order \cite{Zhao:2019aa, PhysRevB.99.195110, PhysRevResearch.2.033459} transitions or \textit{pseudocriticality}---the latter describes a transition that looks continuous on finite length scales but eventually turns first-order in the thermodynamic limit.
It remains open how all of these phenomena are connected
\cite{PhysRevX.5.041048, PhysRevX.7.031051, PhysRevB.102.201116, PhysRevB.102.020407, takahashi2024so5multicriticalitytwodimensionalquantum}.

\begin{figure}[t]
\includegraphics[width=0.967\linewidth]{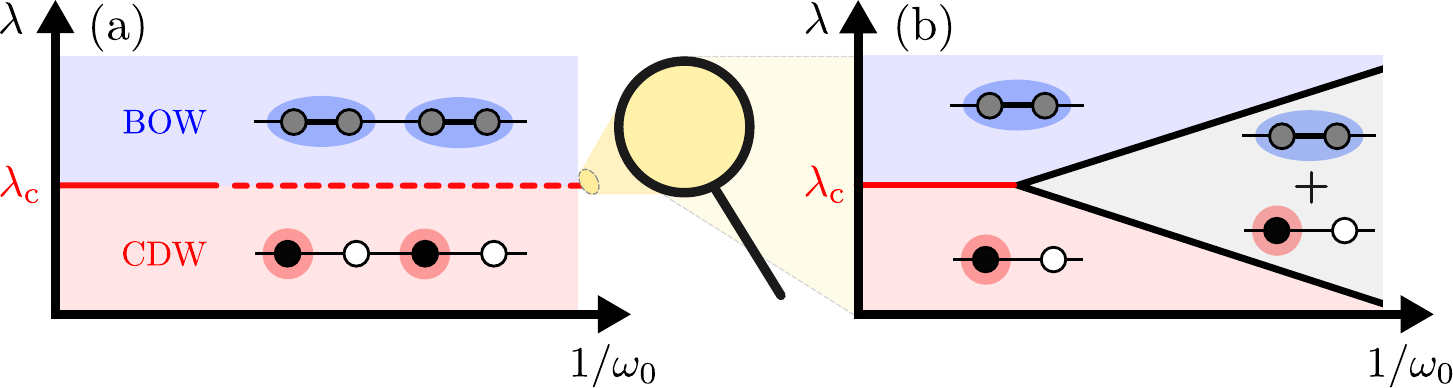}
\caption{%
Schematic illustration of the CDW--BOW transition in the SSH-Holstein model.
(a) In a low-energy Dirac picture, the continuous deconfined transition (solid line) can be tuned first-order (dashed line) via $\omega_0$ while preserving its enhanced U(1) symmetry (red).
(b) On the lattice, the first-order line splits into two continuous transitions enclosing a narrow intermediate phase where both order parameters are finite. U(1) symmetry is only weakly broken on the lattice, so that scenario (a) remains accessible in a large crossover regime giving rise to a symmetry-enhanced pseudo-first-order transition.
}
\label{fig:illu}
\end{figure}

In this Letter, we report how deconfined criticality can 
be tuned towards the conceptual counterpart of pseudocriticality, \ie,
a \textit{pseudo-first-order transition} that clearly looks discontinuous on intermediate length scales but is not in the thermodynamic limit.
To this end, we consider a one-dimensional (1D) realization of deconfined criticality \cite{PhysRevB.99.075103, PhysRevB.99.165143, PhysRevB.99.205153, PhysRevB.100.125137},
which has the advantage that the critical theory is known analytically via Luttinger-liquid theory
and that powerful computational methods allow for large-scale simulations.
A continuous transition can occur between 
charge-density-wave (CDW) and bond-order-wave (BOW) phases
[for illustrations see Fig.~\ref{fig:illu}(a)]
which break bond- and site-inversion symmetries, respectively. 
Originally discovered
in the 1D extended Hubbard model \cite{PhysRevLett.92.236401}, 
its tunability is complicated 
because the narrow BOW phase only appears on large system sizes and strong couplings are required to destroy the continuous transition.

Here we identify competing electron-phonon couplings as an ideal setup to generate a tunable transition.
CDW and BOW orders naturally occur from a coupling to Holstein \cite{HOLSTEIN1959325} and Su-Schrieffer-Heeger (SSH) \cite{PhysRevLett.42.1698} phonons via the Peierls instability where a periodic lattice distortion modulates the charge and bond density, respectively. In the adiabatic phonon limit, each of these couplings generates a static mean-field-like mass term in a (1+1)D Dirac theory, whereas a finite phonon frequency $\omega_0 > 0$ allows to systematically include quantum fluctuations to the phonon order parameter.
Individually, Holstein and SSH couplings have been studied in great detail \cite{2018EPJB...91..204H}, but
the combination of both has only revealed weak signatures of a continuous transition for small lattice sizes
\cite{PhysRevLett.117.206404}. Here we provide strong evidence for a deconfined CDW--BOW transition with an emergent chiral U(1) symmetry using large-scale quantum Monte Carlo (QMC) simulations \cite{PhysRevLett.119.097401}. Tuning $\omega_0 \to 0$ suppresses the quantum fluctuations 
that stabilize the deconfined critical point and unravels a pseudo-first-order transition
with enhanced U(1) symmetry, 
as depicted in Fig.~\ref{fig:illu}. 
As this symmetry is only weakly broken on the lattice,
we find a broad crossover regime in which this exotic phenomenon occurs and even exhibits scaling behavior.
Our findings are relevant for a large class of quantum systems in which symmetry
 enhancement plays an important role.

\textit{Model.}---%
We consider the 1D SSH-Holstein model $\hat{H} = \hat{H}_\mathrm{el} + \hat{H}_\mathrm{ph}$.
The electronic part is given by
\begin{align}
\label{eq:Hel}
\hat{H}_\mathrm{el}
	=
	 \sum_{i} \bigl(-t + \gb \, \Q{i,\mathrm{b}} \bigr) \hat{B}_i
	+ g_\mathrm{s} \sum_{i} \Q{i,\mathrm{s}} \left(\fden{i} - 1\right)
\end{align}
where $\hat{n}_i  =  \sum_\sigma \hat{c}^\dagger_{i,\sigma} \hat{c}_{i,\sigma}$ and $\hat{B}_i = \sum_\sigma (\hat{c}^\dagger_{i,\sigma} \hat{c}_{i+1,\sigma} + \Hc)$ define the local site and bond density operators
and $\hat{c}_{i,\sigma}$ annihilates a fermion with spin $\sigma$ at site $i\in\{1,\dots, L\}$. 
$\hat{H}_\mathrm{el}$
contains the nearest-neighbor hopping of electrons on a 1D chain 
and their interaction with site/bond phonons via the couplings $g_\mathrm{s/b}$. The phonon Hamiltonian,
\begin{align}
\hat{H}_\mathrm{ph}
	=
	\sum_i \sum_{\alpha\in\{\mathrm{s},\mathrm{b}\}} \left( \frac{1}{2M_\alpha} \P{i,\alpha}^2 + \frac{K_\alpha}{2} \Q{i,\alpha}^2 \right) \, ,
\end{align}
describes local harmonic oscillators in terms of their displacements $\Q{i,\alpha}$ and momenta $\P{i,\alpha}$ which are defined on the sites/bonds ($\mathrm{s/b}$) of our chain. We choose the optical phonon frequencies $\omega_0 = \sqrt{K_\alpha / M_\alpha}$ to be equal for both $\alpha$ and define the dimensionless couplings $\lambda_\mathrm{s} = g_\mathrm{s}^2/(4K_\mathrm{s}t)$ and $\lambda_\mathrm{b} = g_\mathrm{b}^2/(K_\mathrm{b}t)$. We consider half-filling, use periodic boundary conditions, choose $t$ as the unit of energy, and inverse temperatures $\beta t = 4L$.

\textit{Methods.}---%
For $\omega_0 \to 0$ and $L, \beta\to \infty$, we perform an exact mean-field analysis (see below).
For $\omega_0 > 0$ and/or finite $L,\beta$
we use an exact QMC method that samples a 
perturbation expansion of the partition function
in terms of world-line configurations. 
Our method makes use of the global directed-loop updates \cite{PhysRevB.59.R14157, PhysRevE.66.046701} and their generalization to retarded interactions \cite{PhysRevLett.119.097401, AppendixSSHHS} to overcome the autocorrelation problem that has been detrimental for direct phonon sampling in QMC methods
\cite{PhysRevB.75.245103, Hohenadler2008}.

\textit{Exact ground state in the adiabatic phonon limit.---}%
The Peierls instabilities
can be understood
in the limit
$\omega_0 \to 0$, \ie, $M_\alpha \to \infty$ at fixed $K_\alpha$. Then, the phonon momenta $\P{i,\alpha}$ drop out of the Hamiltonian such that $\Q{i,\alpha} \to Q_{i,\alpha}$ become classical variables which need to be minimized to determine the ground state.
For Holstein phonons,
the exact solution is given by the staggered displacements
$Q_{i,\mathrm{s}}= (-1)^i \Delta_\mathrm{CDW} / g_\mathrm{s}$ which induce CDW order for any $\lambda_\mathrm{s}>0$,
whereas SSH phonons lead to BOW order for any $\lambda_\mathrm{b}>0$ via
$Q_{i,\mathrm{b}} = - t_\mathrm{b} / g_\mathrm{b} + (-1)^i \Delta_\mathrm{BOW} / (2 g_\mathrm{b})$ \footnote{Note that our optical bond phonons only differ from the original acoustic SSH phonons \cite{PhysRevLett.42.1698} in an additional renormalization of the electron hopping via $t_\mathrm{b}$.}.
For $\lambda_\mathrm{s}, \lambda_\mathrm{b}>0$,
the mean-field equations for $\Delta_\mathrm{CDW}$, $\Delta_\mathrm{BOW}$, and $t_\mathrm{b}$ are solved self-consistently. 
We obtain
$\hat{H}_\mathrm{el} = \frac{L}{2\pi} \int_0^\pi dk \sum_{\sigma} \mathbf{\hat{c}^\dagger}_{k,\sigma} \mathcal{H}_k \, \mathbf{\hat{c}}_{k,\sigma}$ where
\begin{align}
\label{eq:Hsp}
\mathcal{H}_k = \tilde{\epsilon}(k) \, \tau_z + \Delta_\mathrm{CDW} \, \tau_x - \Delta_\mathrm{BOW} \sin(k) \, \tau_y \, .
\end{align}
Here, $\tilde\epsilon(k) = - 2\left(t + t_\mathrm{b}\right) \cos(k)$ is the renormalized free band dispersion and $\tau_i$ are the Pauli matrices which act on $\mathbf{\hat{c}^\dagger}_{k,\sigma} = ({\hat{c}^\dagger}_{k,\sigma}, {\hat{c}^\dagger}_{k+\pi,\sigma})$.
The eigenvalues of Eq.~\eqref{eq:Hsp} are
$
\lambda_{k \pm} = \pm \, [{ \tilde\epsilon(k)^2 + \Delta_\mathrm{CDW}^2 + \Delta_\mathrm{BOW}^2 \sin^2(k) }]^{1/2}
$
and the phonon potential becomes
\begin{align}
\label{eq:Hph2}
H_\mathrm{ph}
	=
	\frac{L}{8t} \left[ \frac{\Delta_\mathrm{CDW}^2}{\lambda_\mathrm{s}} + \frac{\Delta_\mathrm{BOW}^2}{\lambda_\mathrm{b}}  + \frac{4 t_\mathrm{b}^2}{\lambda_\mathrm{b}}\right] \, .
\end{align}

If we expand  Eq.~\eqref{eq:Hsp} around the Fermi points at $\kF = \pm\pi/2$,
$\mathcal{H}_k$ can be understood as a (1+1)D Dirac system with anticommuting masses \cite{PhysRevB.80.205319} that can be combined into a vector $\mathbf{\Delta} = (\Delta_\mathrm{CDW}, \Delta_\mathrm{BOW})$ and lead to a gap $|\mathbf{\Delta}|$ in the energy dispersion
$\lambda_{p\pm} = \pm \sqrt{ (\tilde{v}_\mathrm{F} p)^2 + |\mathbf{\Delta}|^2}$.
For $\lambda_\mathrm{s} = \lambda_\mathrm{b}$, both mass terms exhibit the same potential \eqref{eq:Hph2}, \ie, $H_\mathrm{ph}\propto |\mathbf{\Delta}|^2$, such that our Dirac system displays a chiral U(1) symmetry with $\mathbf{\Delta} = |\mathbf{\Delta}| (\cos \theta, \sin \theta)$ which allows for arbitrary rotations $\theta \in [0,2\pi)$ between the two orders. This symmetry has been discussed, \eg, in the context of solitons with irrational quantum numbers \cite{PhysRevLett.47.986, Fradkin_2013}. However, for $\lambda_\mathrm{s} > \lambda_\mathrm{b}$ ($\lambda_\mathrm{s} < \lambda_\mathrm{b}$) the phonon potential \eqref{eq:Hph2} pins the phase at $\theta\in\{0,\pi\}$ ($\theta \in \{\pi/2, 3\pi/2\}$) and the electrons exhibit only CDW (BOW) order. As a result, the linearized theory predicts a direct first-order CDW--BOW transition with enhanced U(1) symmetry at $\lambda_\mathrm{s} = \lambda_\mathrm{b}$ described by a discontinuity fixed point \cite{PhysRevLett.35.477, PhysRevB.26.2507}.

\begin{figure}[t]
\includegraphics[width=0.97\linewidth]{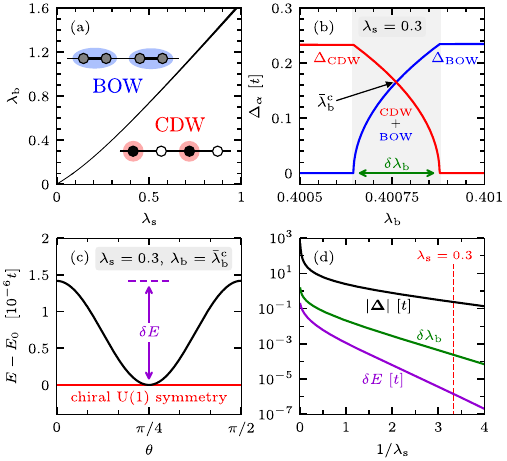}
\caption{%
Ground-state properties for $\omega_0 = 0$. (a) Phase diagram as a function of the couplings $\lambda_{\mathrm{s}}$ and $\lambda_\mathrm{b}$.
The width of the phase boundary encodes the extent of the intermediate phase.
(b) Order parameters $\Delta_\mathrm{CDW/BOW}$ for $\lambda_\mathrm{s}=0.3$ across the mixed phase of width $\delta \lambda_\mathrm{b}$.
Here $\bar\lambda_\mathrm{b}^\mathrm{c}$ defines the coupling where $\Delta_\mathrm{CDW} = \Delta_\mathrm{BOW}$.
(c) Modulation of the total energy per site if the angle
 $\theta = \arctan(\Delta_\mathrm{CDW}/\Delta_\mathrm{BOW})$ between order parameters is varied at fixed $\lambda_\mathrm{b} = \bar\lambda_\mathrm{b}^\mathrm{c}$. Coexistence reduces the energy by $\delta E$ whereas an exact U(1) symmetry requires $\delta E = 0$ for all $\theta$.
(d) Single-particle gap $|\mathbf{\Delta}|$ and energy gain $\delta E$ at $\bar\lambda_\mathrm{b}^\mathrm{c}$ as well as $\delta\lambda_\mathrm{b}$ as a function of $1/\lambda_\mathrm{s}$.
}
\label{fig:MF}
\end{figure}

Taking into account the nonlinear band dispersion in Eq.~\eqref{eq:Hsp} leads to important changes
to the phase diagram shown in Fig.~\ref{fig:MF}(a): The CDW--BOW phase boundary is not only tilted but also split into a narrow intermediate phase \cite{PhysRevB.28.2653} where site \textit{and} bond inversion are broken, thus indicating a ferroelectric phase \cite{Rossler1990, 2024arXiv240611952Z}.
The evolution of the two mass terms across the intermediate regime is displayed in Fig.~\ref{fig:MF}(b) for $\lambda_\mathrm{s} = 0.3$ 
where the width of the mixed phase is only 
$\delta\lambda_\mathrm{b} \approx 0.00023$. 
The angle $\theta$ between CDW and BOW order parameters 
evolves continuously from $0$ to $\pi/2$ signaling the absence of the chiral U(1) symmetry.
However, the energy splitting $\delta E /t$ 
of the chiral manifold is only $\mathcal{O}(10^{-6})$ 
at $\theta = \pi / 4$, 
as shown in Fig.~\ref{fig:MF}(c), which is significantly smaller than the gap $|\mathbf{\Delta}|/t \approx0.234$ and indicates a clear separation of energy scales. 
Figure \ref{fig:MF}(d) shows that this separation of scales occurs for all couplings
and that $|\mathbf{\Delta}|$, $\delta\lambda_\mathrm{b}$, and $\delta E$ exhibit
$\exp(-a/\lambda_\mathrm{s})$ behavior for $\lambda_\mathrm{s} \lesssim 1$.
The CDW+BOW state is separated from the clean phases by two second-order phase transitions that exhbit mean-field critical exponents; \eg, we find
$\beta=1/2$ in Fig.~\ref{fig:MF}(b).

\begin{figure}[t]
\includegraphics[width=\linewidth]{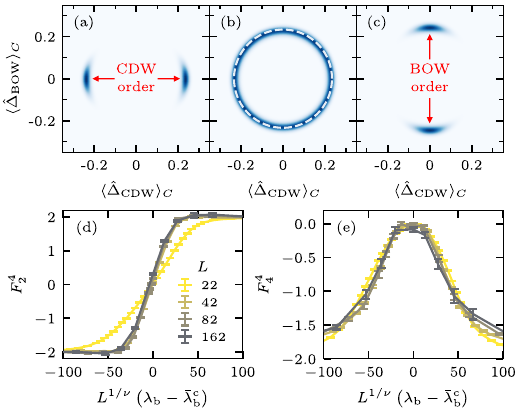}
\caption{%
Symmetry-enhanced pseudo-first-order transition at $\omega_0 = 0$ and $\lambda_\mathrm{s}=0.3$.
(a)--(c) Histograms of the order parameter $(\langle\hat\Delta_\mathrm{CDW}\rangle_C, \langle\hat\Delta_\mathrm{BOW}\rangle_C)$ for (a) $\lambda_\mathrm{b} = 0.39$, (b) $\lambda_\mathrm{b} = 0.4005$, and (c) $\lambda_\mathrm{b} = 0.41$ at $L=82$, $\beta t = 4L$. The dashed white circle in (b) uses the mean-field gap $|\mathbf{\Delta}|/t \approx 0.234$. (d),(e) Data collapse of the integrated measures $F_2^4$ and $F_4^4$ using $1/\nu = 2$.
}
\label{fig:symmetry_om0}
\end{figure}

\textit{Pseudo-first-order transition with enhanced U(1) symmetry.}---%
Although the chiral U(1) symmetry is lifted by
our lattice model, 
the clear separation of energy scales in Fig.~\ref{fig:MF}(d) suggests that it remains visible in a broad crossover regime of intermediate energy and length scales. 
To test this hypothesis, we use our QMC method to calculate histograms of the phonon order parameters
\cite{AppendixSSHHS}
\begin{align}
\label{eq: orderp}
\hat{\Delta}_\mathrm{CDW/BOW}
 = \frac{f_\mathrm{s/b}}{L\beta} \int_0^\beta d\tau \sum_i (-1)^i  \,
\Q{i,\mathrm{s/b}}(\tau)
\end{align}
where $f_\mathrm{s} = g_\mathrm{s}$ and $f_\mathrm{b} = 2g_\mathrm{b}$ to match the mean-field gaps for $\omega_0 = 0$ and $L, \beta \to \infty$.
For $\omega_0 = 0$ and $\lambda_\mathrm{s} =0.3$, the combined histograms $(\langle\hat\Delta_\mathrm{CDW}\rangle_C, \langle\hat\Delta_\mathrm{BOW}\rangle_C)$---here $\langle \bullet \rangle_C$ denotes the expectation value for a single Monte Carlo configuration---are depicted in Figs.~\ref{fig:symmetry_om0}(a)--(c) across the CDW--BOW interface for $L=82$ and $\beta t = 4L$. In the CDW/BOW phases,
we find two peaks representative of their $\mathds{Z}_2$ order parameters. 
At the transition, the latter merge into a perfect circle of radius $|\mathbf{\Delta}|$
which is in good agreement with the mean-field prediction. Together with the uniform angle dependence of its intensity, this is a clear signature of the enhanced U(1) symmetry expected from the low-energy Dirac theory.

\begin{figure}[t]
\includegraphics[width=\linewidth]{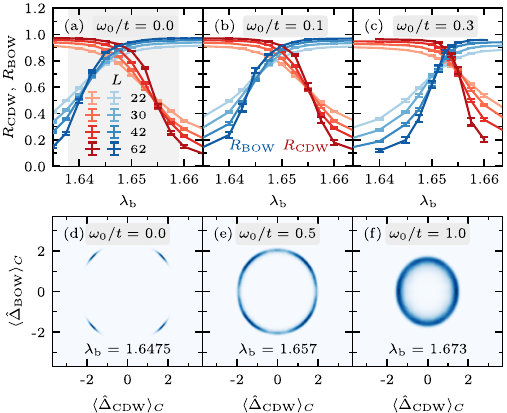}
\caption{%
Effects of quantum lattice fluctuations on the intermediate phase at $\lambda_\mathrm{s}=1.0$.
(a)--(c) Finite-size dependence of the correlation ratios $R_\mathrm{CDW/BOW}$ for different $\omega_0$.
For $\omega_0 = 0$ the shaded area indicates the exact mean-field result for the intermediate phase.
(d)--(f) Histograms of the order parameter in the mixed phase for different $\omega_0$. Here $L=62$, $\beta t = 4L$.
}
\label{fig:stab}
\end{figure}

The emergent U(1) symmetry can also be detected via the integrated measures \cite{PhysRevLett.115.267203, PhysRevB.99.195110}
$F_\ell^a = \langle r^a \cos(\ell \theta) \rangle$
where $(\bar{\Delta}_\mathrm{CDW}, \bar{\Delta}_\mathrm{BOW}) = r \left(\cos \theta, \sin\theta\right)$
and $\bar{\Delta}_\alpha = {\hat{\Delta}_\alpha}/{{\langle\hat{\Delta}_\alpha^2\rangle^{1/2}}}$.
In particular, the cumulants
$
F_2^4 = \expv{\bar{\Delta}_\mathrm{CDW}^4 - \bar{\Delta}_\mathrm{BOW}^4}$
and
$
F_4^4 = \expv{\bar{\Delta}_\mathrm{CDW}^4 - 6 \bar{\Delta}_\mathrm{CDW}^2 \bar{\Delta}_\mathrm{BOW}^2 + \bar{\Delta}_\mathrm{BOW}^4}$
have a simple form accessible within our QMC simulations. Enhanced U(1) symmetry requires $F^a_\ell \to 0$ which is confirmed in Figs.~\ref{fig:symmetry_om0}(d),(e) for $F_2^4$ and $F_4^4$. Moreover, these cumulants
fulfil the scaling relations
$F^a_\ell = f^a_\ell \boldsymbol{(}  L^{1/\nu} (\lambda_\mathrm{b} - \lambda_\mathrm{b}^\mathrm{c}) \boldsymbol{)}$.
For a discontinuity fixed point, $1/\nu = d + 1$ is determined by the space-time dimension \cite{PhysRevLett.35.477, PhysRevB.26.2507}.
Setting $1/\nu =2$, Figs.~\ref{fig:symmetry_om0}(d),(e) show excellent data collapses for $L\gtrsim 42$.
Note that $1/\nu = 2$ is also the mean-field exponent for the two transitions into the intermediate phase. However, the mixed state requires 
the breakdown of the U(1) symmetry [\cf Fig.~\ref{fig:stab}(d)] for which $F_4^4 \to -4$. Hence, at $\lambda_\mathrm{s} =0.3$ our system is still far away from crossing over towards the real ground state.
On the other side, for $L=22$ in Figs.~\ref{fig:symmetry_om0}(d),(e) we have not yet reached proximity to the discontinuity fixed point.

\begin{figure}[t]
\includegraphics[width=\linewidth]{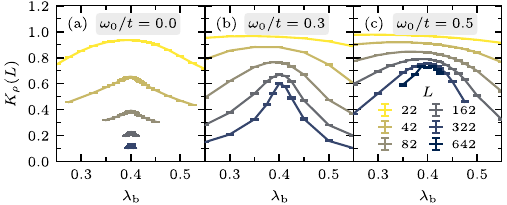}
\caption{%
Quantum-lattice-fluctuation effects at fixed $\lambda_\mathrm{s}=0.3$.
Finite-size analysis of the Luttinger parameter $K_\rho(L)$ for (a) $\omega_0/t = 0.0$, (b) $\omega_0/t = 0.3$, and (c) $\omega_0/t = 0.5$ using $\beta t = 4 L$.
}
\label{fig:Krho}
\end{figure}

\textit{Deconfined quantum criticality.}---%
Tuning the phonon frequency away from $\omega_0 = 0$ allows us to include quantum fluctuations in a systematic way.
To investigate their effect on the CDW--BOW transition region, we calculate
$K_\rho(L) = \pi \, S_\rho(q_1)/q_1 $
from the static charge structure factor at small momentum transfer $q_1 = 2\pi/L$ \cite{PhysRevLett.92.236401}.
For $L\to\infty$, it converges to the charge Luttinger parameter $K_\rho$
which remains finite if the charge sector is gapless and scales to zero otherwise.
Figure \ref{fig:Krho} shows $K_\rho(L)$ for fixed $\lambda_\mathrm{s} = 0.3$ but different $\omega_0$.
At $\omega_0 = 0$ the charge gap is finite for all $\lambda_\mathrm{b}$ such that $K_\rho(L) \to 0$.
As we increase $\omega_0$, a peak develops at the CDW--BOW interface and converges to  $K_\rho > 0$ signalling the closing of the charge gap 
beyond a critical $\omega_0$. Note that previous QMC simulations only reached $L\leq 42$ for the parameters presented in Fig.~\ref{fig:Krho}(c) \cite{PhysRevLett.117.206404}, for which we estimate $K_\rho \approx 0.7$.

In 1D, gapless states of matter are usually described by Luttinger-liquid theory \cite{10.1093/acprof:oso/9780198525004.001.0001}.
In our case, spin excitations are always gapped, so we only need to consider a sine-Gordon model for the charge degrees of freedom,
$
\mathcal{H}_\rho
	= \frac{v_\rho}{2}  \big[ K_\rho (\partial_x \vartheta_\rho)^2 + K^{-1}_\rho (\partial_x \phi_\rho)^2 + \lambda_\rho \cos(\sqrt{8\pi} \phi_\rho) \big]
$.
For $K_\rho \geq 1$ the cosine term is irrelevant, leaving us with a Gaussian theory with gapless charge excitations. 
However, it turns relevant for $K_\rho < 1$; then
$\phi_\rho$ gets pinned at different minima
for $\lambda_\rho > 0$ ($\lambda_\rho < 0$) 
inducing
CDW (BOW) order with a finite charge gap. 
At the deconfined CDW--BOW transition
we tune through $\lambda_\rho = 0$
where the cosine term disappears and leads to a Gaussian critical point with zero charge gap.
The latter exists for $1/4 < K_\rho < 1$ until $8\kF$ Umklapp scattering becomes relevant  \cite{PhysRevB.25.4925, PhysRevLett.92.236401}.

\begin{figure}[t]
\includegraphics[width=\linewidth]{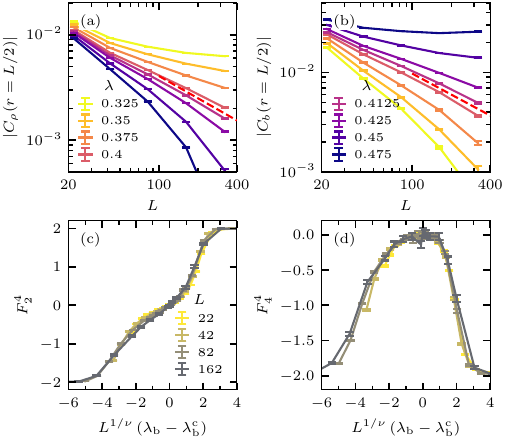}
\caption{%
Signatures of 1D deconfined criticality at the CDW--BOW transition. (a),(b) Charge/bond correlations at maximum distance, $C_\mathrm{\rho/b}(r=L/2)$, as a function of $L$. The dashed red line indicates the $L^{-K_\rho}$ decay at the critical point where $K_\rho \approx 0.7$.
(c),(d) Data collapse of $F_2^4$,$F_4^4$ using $1/\nu = 2-2K_\rho$.
Here $\omega_0/t =0.5$, $\lambda_\mathrm{s}=0.3$, and $\beta t = 4L$.
}
\label{fig:CDWBOW}
\end{figure}
Figure \ref{fig:CDWBOW} provides further evidence for a deconfined critical point at $\omega_0/t =0.5$ and $\lambda_\mathrm{s}=0.3$. The charge/bond correlation functions $C_{\rho / b}(r) = \sum_i \langle \tilde{O}^{\mathrm{s/b}}_i \tilde{O}^{\mathrm{s/b}}_{i+r}  \rangle / L$---here $\tilde{O}_i^\alpha = \hat{O}_i^\alpha -\langle  \hat{O}_i^\alpha \rangle$ and $\hat{O}_i^\alpha\in\{\hat{n}_i, \hat{B}_i\}$---decay with the same power law $r^{-K_\rho}$ only at the critical point, as shown in Figs.~\ref{fig:CDWBOW}(a),(b), whereas $|C_{\rho/b}(r\to\infty)| \to const$ and $C_{b/\rho}(r\to\infty)\to 0$ in the CDW/BOW phase. The unpinning of the charge field $\phi_\rho$ in $\mathcal{H}_\rho$
leads to an emergent chiral U(1) symmetry at criticality, as indicated by $F_2^4, F_4^4 \to 0$ in Figs.~\ref{fig:CDWBOW}(c),(d). A data collapse of the latter is in good agreement with $1/\nu = 2- 2K_\rho$ predicted by bosonization \cite{PhysRevLett.92.236401}. Because CDW order is weak for $\lambda_\mathrm{s} = 0.3$, we still observe a finite-size drift for $\lambda_\mathrm{b} < \lambda_\mathrm{b}^\mathrm{c}$.

\textit{Stability of the mixed phase.}---%
The effects of quantum lattice fluctuations are demonstrated in Fig.~\ref{fig:stab} for $\lambda_\mathrm{s}=1.0$.
We estimate the extent of the intermediate phase using the correlation ratios
$R_\mathrm{CDW/BOW} = 1 - S_{\rho/b}(q=\pi + q_1) / S_{\rho/b}(q=\pi)$ which scale to one (zero) in the ordered (disordered) phase and become scale invariant at criticality. For $\omega_0 = 0$ in Fig.~\ref{fig:stab}(a), we find an extended intermediate regime where both $R_\mathrm{CDW/BOW}(L) \to 1$;
however, available systems are still too small to converge to the exact mean-field prediction. This phase remains stable for small but finite $\omega_0$ but its width decreases, \cf Figs.~\ref{fig:stab}(b),(c). Histograms of the phonon order parameter are presented in Figs.~\ref{fig:stab}(d)--(f). At $\omega_0 = 0$ we find peaks at finite angles $\theta$ characteristic for the intermediate phase, but increasing $\omega_0$ at a fixed length scale $L=62$ gradually restores the U(1) symmetry and reduces its radius towards a point when approaching the deconfined critical point.
Hence, pseudo-first-order behavior with its separation of energy scales $0 < \delta E \ll | \mathbf{\Delta}|$ also emerges from the deconfined critical point at strong couplings.

\textit{Discussion.}---%
Dirac systems with anticommuting mass terms are a generic setup to generate emergent symmetries and compatible defects that can drive a deconfined phase transition \cite{PhysRevB.80.205319}. Here we have shown that competing electron-phonon interactions are an ideal realization for this because the limit $\omega_0 \to 0$
represents an exactly-solvable Dirac model and, on top of this, quantum fluctuations can be added systematically via the retardation range; 
such tunability has been missing
 in previous realizations of retardation-induced 1D deconfined criticality \cite{PhysRevResearch.2.023013, 2024arXiv241213263W}.
 Chiral U(1) symmetry is inherent to Luttinger-liquid theory but also appears at our pseudo-first-order transition. 
 Therefore, we have found a consistent setup in which first- and second-order transitions with enhanced symmetries can be tuned into each other using large-scale numerics, something that is desired for 2D quantum magnets \cite{takahashi2024so5multicriticalitytwodimensionalquantum}.
As we only find a pseudo-first-order transition with weak symmetry breaking,
it is natural to ask if this is also the case for analogous transitions found in the context of (2+1)D deconfined criticality \cite{Zhao:2019aa, PhysRevB.99.195110, PhysRevResearch.2.033459}.
 We expect that our mechanism also applies to higher-dimensional Dirac systems where enhanced symmetries can be weakly broken \cite{PhysRevB.80.205319, PhysRevB.82.035429}.
 In (2+1)D Dirac systems, previously observed signatures of deconfined criticality \cite{PhysRevLett.119.197203} might only exist in a crossover regime towards a narrow coexistence phase \cite{PhysRevResearch.2.022005}.
Symmetry enhancement at first-order transitions also plays an important role in the SO(5) theory of high-$T_\mathrm{c}$ superconductivity \cite{doi:10.1126/science.275.5303.1089, RevModPhys.76.909} but, as in classical systems \cite{PELISSETTO2002549},
is likely to be unstable \cite{PhysRevB.67.054505}.

Our results suggest that the continuous CDW--BOW transition in the 1D SSH-Holstein model splits into two transitions once $K_\rho < 1/4$. It is natural to ask if the same could happen in the 1D extended Hubbard model where the CDW--BOW transition is assumed to turn first-order beyond a tricritical point \cite{PhysRevLett.92.236401}.
The emergence of a mixed phase \cite{PhysRevB.99.075103, PhysRevB.99.165143} and symmetry enhancement \cite{PhysRevB.108.245152} have also been observed in other 1D systems. 
In principle, different scenarios are possible for the disappearance of deconfined criticality;
our electron-phonon model with its exactly solvable limit is an ideal starting point for future studies.

It is remarkable that clear signatures of a first-order transition
can persist over a wide range of system sizes but not describe the true ground state. 
Thus, our pseudo-first-order transition belongs to the same category of crossover phenomena as pseudocriticality; 
interestingly, both exhibit properties of first- and second-order transitions. 
Few cases have been reported for classical systems 
without symmetry enhancement \cite{PhysRevE.62.3780, Schreiber_2005, LUNDOW2011120, PhysRevLett.108.045702}; the free-energy barrier between coexisting states leads to different finite-size effects like negative peaks in Binder ratios absent in our case.
Symmetry-enhanced pseudo-first-order behavior should also occur in classical anisotropic O(2) models where cubic anisotropy can act as a dangerously irrelevant perturbation to a direct spin-flop transition \cite{PhysRevB.11.478, PhysRevB.15.3510}.

\textit{Summary.}---%
Using exact QMC simulations, we showed that the CDW--BOW transition in the 1D SSH-Holstein model can be tuned from second- to pseudo-first-order. At both transitions we found an enhanced U(1) symmetry and scaling behavior. In the limit $\omega_0 \to 0$, an exact solution revealed that the pseudo-first-order transition is a slow crossover towards a sequence of continuous transitions. 
Increasing $\omega_0$ reduces the width of the enclosed intermediate phase, gradually restores U(1) symmetry, and tunes our system back to the deconfined critical point.
Our finding of a symmetry-enhanced pseudo-first-order transition is relevant far beyond competing Peierls instabilities, as emergent symmetries naturally occur in Dirac systems but also in other contexts like superconductivity, frustrated magnetism, or exotic criticality.

\begin{acknowledgments}
\textit{Acknowledgments.}---%
I acknowledge helpful discussions with Fakher Assaad, Pedro C\^onsoli, Igor Herbut, Lukas Janssen, Daniel Lozano G\'omez, and Matthias Vojta.
This work was supported by the Deutsche Forschungsgemeinschaft
through the W\"urzburg-Dresden Cluster of Excellence on Complexity and Topology
in Quantum Matter---\textit{ct.qmat} (EXC 2147, Project No. 390858490).
I gratefully acknowledge the computing time made available to me on the high-performance computer at the NHR Center of TU Dresden. This center is jointly supported by the Federal Ministry of Education and Research and the state governments participating in the NHR \footnote{\url{https://www.nhr-verein.de/unsere-partner}}.
\end{acknowledgments}

\appendix

\section{Appendix}

\textit{Retarded interactions.}---%
As the phonons appear only quadratically in $\hat{H}$, they can be integrated out exactly using the coherent-state path integral.
The resulting partition function
$Z = Z_\mathrm{ph}^0 \int \fcohmeasure \, e^{- \mathcal{S}[\fcohcr{}, \fcohan{}]}$ only depends on the electronic action $\S = \S_t + \S_\mathrm{ret}$ where
\begin{align}
\S_\mathrm{ret}
	=
	- \iint_0^\beta d\tau d\tau' \sum_{i \alpha}
	O_{i}^{\alpha}(\tau) \, w_\alpha \, D(\tau - \tau') \, O_{i}^{\alpha}(\tau')
\end{align}
is nonlocal in imaginary time and mediated by the free-phonon propagator $D(\tau) = \omega_0 \, e^{-\omega_0 \tau} / (1-e^{-\beta \omega_0})$. 
For Holstein phonons, $O_i^\mathrm{s} = n_i - 1$ and $w_\mathrm{s} = 2\lambda_\mathrm{s} t$, whereas for SSH phonons $O_i^\mathrm{b} = B_i$ and $w_\mathrm{b} = \lambda_\mathrm{b} t/2$.
For $\omega_0 \to 0$, the interaction range $D(\tau)\to 1/\beta$ becomes infinite,
justifying an exact mean-field solution for $\beta\to\infty$, whereas for $\omega_0 > 0$ quantum fluctuations can be included in a controlled manner. Our QMC method is based on a perturbation expansion of $Z$ in the full action $\mathcal{S}$, as described in Ref.~\cite{PhysRevLett.119.097401}.

\textit{Equivalence of electron and phonon susceptibilities.}---%
For electron-phonon models with a quadratic phonon potential and a linear coupling to the displacement operators, \ie, the type of models considered in this work,
 the phonon correlation functions are fully determined by higher-order correlations of the electronic operators they couple to \cite{PhysRevB.94.245138}. In the following, we derive how charge and phonon order parameters are related.

For Holstein-type electron-phonon couplings, we consider the charge and phonon order parameters
\begin{gather}
\hat{m}_\mathrm{CDW}
	=
	\frac{1}{L\beta}
	\int_0^\beta d\tau \sum_i (-1)^i \left[ \fden{i}(\tau) - 1 \right] \, ,
\\
\hat\Delta_\mathrm{CDW}
	= \frac{ g_\mathrm{s}}{L \beta} \int_0^\beta d \tau \sum_i (-1)^i \, \Q{i,\mathrm{s}}(\tau) \, .
\label{eq:SM_order}
\end{gather}
We include the electron-phonon coupling $g_\mathrm{s}$ into the definition of $\hat\Delta_\mathrm{CDW}$ because then the final result can be expressed by only using $\lambda_\mathrm{s}$. Within the path-integral formalism, the phonon fields can be derived from generating functionals. After integrating out the phonons, the source terms appear in the retarded electronic action, from which we can derive the equivalent electronic observables. This procedure has been described in detail in Ref.~\cite{PhysRevB.94.245138}. For example, the expectation value of a single displacement operator becomes
\begin{align}
\label{eq:SM_1stQ}
\langle \Q{i,\mathrm{s}}(\tau) \rangle
	= 
	- 2 \sqrt{\frac{\lambda_\mathrm{s} t }{K_\mathrm{s}}}
	\int_0^\beta d\tau' \, D_+(\tau - \tau') \, \langle\fden{i}(\tau') - 1\rangle
\end{align}
where $D_+(\tau) = [D(\tau)+D(\beta - \tau)]/2$ is the symmetrized phonon propagator. Because $\int_0^\beta d\tau \, D_+(\tau-\tau')=1$, the integrals in Eq.~\eqref{eq:SM_order} drop out and leave no explicit dependence on the phonon frequency $\omega_0$. We obtain the following relations between the cumulants required to calculate $F_2^4$ and $F_4^4$:
\begin{gather}
\label{eq:SM_1st}
\langle\hat\Delta_\mathrm{CDW}\rangle
	= - 4 \lambda_\mathrm{s} t \,  \langle\hat{m}_\mathrm{CDW}\rangle \, ,
	\\
\langle \hat{\Delta}_\mathrm{CDW}^2 \rangle
	=
	\left(4 \lambda_\mathrm{s} t\right)^2 \langle \hat{m}^2_\mathrm{CDW} \rangle
	+ \frac{4 \lambda_\mathrm{s} t}{L \beta} \, ,
	\\
\langle \hat{\Delta}_\mathrm{CDW}^4 \rangle
	=
	 \left(4 \lambda_\mathrm{s} t\right)^4 \langle \hat{m}^4_\mathrm{CDW} \rangle
	+ 6 \frac{\left(4\lambda_\mathrm{s} t\right)^3}{L\beta} \langle \hat{m}^2_\mathrm{CDW}\rangle \nonumber\\
	+ 3 \frac{\left(4\lambda_\mathrm{s} t\right)^2}{(L\beta)^2} \, .
\end{gather}
While Eq.~\eqref{eq:SM_1st} follows directly from Eq.~\eqref{eq:SM_1stQ}, the remaining two contain additional terms because after integrating out the phonons the source terms also couple to themselves; for details see Ref.~\cite{PhysRevB.94.245138}. However, the corrections are suppressed at large $L$ and $\beta$, so that the leading terms become equivalent up to constant prefactors.

For our SSH-type electron-phonon coupling, we define 
\begin{gather}
\hat{m}_\mathrm{BOW}
	=
	\frac{1}{L\beta}
	\int_0^\beta d\tau \sum_i (-1)^i \, \hat{B}_i(\tau) \, ,
\\
\hat\Delta_\mathrm{BOW}
	= \frac{2 g_\mathrm{b}}{L \beta} \int_0^\beta d \tau \sum_i (-1)^i \, \Q{i,\mathrm{b}}(\tau) \, .
\label{eq:SM_order2}
\end{gather}
By substituting $\lambda_\mathrm{s} \to \lambda_\mathrm{b} / 4$ and $\hat{\Delta}_\mathrm{CDW} \to \hat{\Delta}_\mathrm{BOW} / 2$ in our previous results, we obtain
\begin{gather}
\langle\hat\Delta_\mathrm{BOW}\rangle
	= - 2 \lambda_\mathrm{b} t \,  \langle\hat{m}_\mathrm{BOW}\rangle \, ,
	\\
\langle \hat{\Delta}_\mathrm{BOW}^2 \rangle
	=
	4 \left[ \left( \lambda_\mathrm{b} t\right)^2 \langle \hat{m}^2_\mathrm{BOW} \rangle
	+ \frac{\lambda_\mathrm{b} t}{L \beta} \right] \, ,
	\\
\langle \hat{\Delta}_\mathrm{BOW}^4 \rangle
	=
	16 \Bigg[
	\left(\lambda_\mathrm{b} t\right)^4 \langle \hat{m}^4_\mathrm{BOW} \rangle
	+ 6 \frac{\left(\lambda_\mathrm{b} t\right)^3}{L\beta} \langle \hat{m}^2_\mathrm{BOW}\rangle \nonumber\\
	+ 3 \frac{\left(\lambda_\mathrm{b} t\right)^2}{(L\beta)^2} \Bigg] \, .
\end{gather}
Because the correction terms are suppressed by $L$ and $\beta$, it does not make any visible difference if we calculate $F_2^4$ and $F_4^4$ from the electron or phonon order parameters.

For $\omega_0 \to 0$, the phonon operators $\Q{i,\alpha}(\tau)$ become classical variables and lose their time dependence. Then, the cumulants derived above are the same for the corresponding equal-time phonon observables. However, this is not true anymore for $\omega_0 > 0$ where the phonon propagators do not drop out and expressions become more complicated \cite{PhysRevB.94.245138}.
We have also seen that mean-field theory becomes exact for $\omega_0 \to 0$ and $L,\beta \to \infty$. Then $\langle \hat{\Delta}_\mathrm{\alpha} \rangle \to \Delta_\alpha$ coincides with the corresponding single-particle gap (\ie, half of the two-particle gap). The prefactors in Eqs.~\eqref{eq:SM_order} and \eqref{eq:SM_order2} are chosen in such a way that both gaps have the same size if we tune our low-energy Dirac theory at $\omega_0 = 0$ to the high-symmetry point. Therefore, we observe the emergence of a perfect circle without the need to renormalize the histograms by the variance of the distribution. For $\omega_0 > 0$ this is not true anymore and we observe slight deformations of our circles which can be scaled away, though. 

In our QMC simulations, phonon observables are not directly accessible because the phonons have been integrated out. However, we can recover them from electronic correlation functions \cite{PhysRevB.94.245138}; the time-integrated susceptibilities derived above have particularly simple forms, for which we just have to calculate the corresponding electronic susceptibilities. The charge field $\hat{m}_\mathrm{CDW}$ and its higher moments can be easily recovered from the world-line configurations because it only consists of diagonal operators. By contrast, the BOW field consists of the off-diagonal hopping operator, which needs to be recovered from the distribution of kinetic-energy vertices.


\begin{thebibliography}{57}
\bibitem [{App()}]{AppendixSSHHS}%
  \BibitemOpen
  \href@noop {} {}\bibinfo {note} {See Appendix for details on retarded
  interactions and how we calculate the phonon order parameters.}\BibitemShut
  {Stop}%
\end{thebibliography}
\end{document}